# Quantum teleportation of high-dimensional spatial modes: Towards an image teleporter

Xiaodong Qiu, Haoxu Guo, and Lixiang Chen†

*Department of Physics and Collaborative Innovation Center for Optoelectronic Semiconductors and Efficient Devices, Xiamen University, Xiamen 361005, China*
†*chenlx@xmu.edu.cn*

**Quantum teleportation[1], lying at the heart of a variety of quantum technologies, has inspired a widespread of research activities[2-24], most of which focused on 2-dimensional qubit states. Multilevel systems, qudits, promise to upgrade and inspire new technical developments in high-dimensional Hilbert space[25]. Whereas, for high-dimensional teleportation, it routinely necessitates several ancillary photons in linear optical schemes[21-23]. A fundamental open challenge remains as to whether we can teleport qudit states with bipartite entanglement only. Here we demonstrate such a high-dimensional teleportation using photonic orbital angular momentum (OAM). The so-called "perfect vortices[26]" are exploited both for conducting the prerequisite entanglement concentration to prepare high-dimensional yet maximal OAM entanglement, and for performing faithful high-dimensional Bell sate measurements (HDBSM) based on sum-frequency generation. We experimentally achieve the average teleportation fidelity 0.8788±0.048 for a complete set of 3-dimensional mutually unbiased bases, for instance, conditional on three specific HDBSM results. More importantly, we succeed in realizing the first quantum teleportation of optical images by exploring full transverse spatial entanglement. From the multi-pixel field of view in the teleported images recorded by the ICCD camera, we estimate the effective channel capacity up to $\kappa > 100$. Our scheme holds promise for future high-volume quantum image transmission.**

Since the seminal proposal by Bennett *et al*.[1], quantum teleportation has aroused numerous attentions as its attractive potentials in some advanced and practical quantum information protocols, e.g., realizing quantum repeaters in long-distance quantum communication[27], constructing distributed quantum networks[28], and conducting measurement-based quantum computation[29,30]. Hitherto, quantum teleportation has been reported in different quantum systems, such as flying photons[2-4], nuclear magnetic resonance (NMR)[7], atom ensembles[10-12], solid-state systems[19], and even thermal light source[24]. Recently, teleportation distance has been broadened to a global scale based on satellite-based photonic entanglement[20]. It is noted that most of these previous schemes focused on the two-dimensional qubit systems only.

Recent years have also witnessed a growing interest in the high-dimensional or multi-level quantum systems, qudits, as it can offer higher information capacity, greater resilience to noises and eavesdropping[31,32], and more efficient quantum logic gates[33]. However, these benefits are only conceivable when we are able to efficiently prepare, manipulate, and measure the genuine high-dimensional quantum states. In particular, bipartite high-dimensional Bell state measurement (HDBSM) is essential for high-dimensional dense coding and quantum teleportation. Several attempts have been made, by means of ancillary photonic bits, to perform the 3-dimensional BSM and to demonstrate the qutrit teleportation, which is encoded by the photon path[22,23]. However, since the number of the required ancillary bits increases significantly with the teleportation dimensionality in the linear optical detection schemes[34], the experimental teleportation of an arbitrary high-dimensional quantum states remains challenging.

Here we demonstrate such a high-dimensional teleportation, by exploiting perfect vortices to prepare high-dimensional yet maximally entangled OAM states and by employing sum-frequency generation to perform faithful high-dimensional Bell sate measurements (HDBSM). Furthermore, we report the first quantum teleportation of optical images, and characterize its high-dimensional feature across the multi-pixel field of view in our quantum image teleporter system.

Twisted photons carrying OAM possesses an inherent capacity for constructing a high-dimensional Hilbert space, thus making it a promising candidate for high-dimensional quantum information processing[35]. It was Allen and coworkers[36] that recognized that the Laguerre-Gaussian (LG) modes of helical phase $\exp(i\ell\phi)$ carries a well-defined OAM of $\ell\hbar$. We note that the OAM entanglement of photon pairs produced by spontaneous parametric down-conversion inevitably suffers from the limited spiral bandwidth[37], i.e., lower-order LG modes appears more frequently than higher-order ones. However, the maximal entanglement is prerequisite for a standard quantum teleportation. To overcome this obstacle, we adopt the modified LG modes, the so-called "perfect vortices[26]", to represent OAM eigenstate, $|\ell\rangle$, see the **Methods** for details. As a result, entanglement concentration is implemented naturally, leading to the high-dimensional yet maximal OAM entanglement. In our experimental demonstration, we first realize the quantum teleportation of 3-dimensinal OAM states, qutrits. Assume Alice wishes to teleport to Bob the quantum state of photon $a$ encoded in the OAM basis as, $|\varphi\rangle_a = \alpha_1|-1\rangle_a + \alpha_2|0\rangle_a + \alpha_3|1\rangle_a$, where the complex coefficients $\alpha_1, \alpha_2, \alpha_3$ fulfill $|\alpha_1|^2 + |\alpha_2|^2 + |\alpha_3|^2 = 1$. For teleporting such an arbitrary 3-

dimensional states, Alice and Bob need share the aforementioned bipartite high-dimensional and maximally entangled state (photons $b$ and $c$),

$$|\Psi\rangle_{bc} = \frac{1}{\sqrt{3}}\left(|-1,1\rangle_{bc} + |0,0\rangle_{bc} + |1,-1\rangle_{bc}\right). \tag{1}$$

Then, the whole state of these three photons can be written as,

$$\begin{aligned}|\Psi\rangle_{abc} = \frac{1}{3}\Big\{&|\psi_{00}\rangle_{ab}\left(\alpha_1|-1\rangle_c + \alpha_2|0\rangle_c + \alpha_3|1\rangle_c\right) + \\ &|\psi_{01}\rangle_{ab}\left(\alpha_1\omega^2|-1\rangle_c + \alpha_2|0\rangle_c + \alpha_3\omega|1\rangle_c\right) + \\ &|\psi_{02}\rangle_{ab}\left(\alpha_1\omega|-1\rangle_c + \alpha_2|0\rangle_c + \alpha_3\omega^2|1\rangle_c\right) + \\ &|\psi_{10}\rangle_{ab}\left(\alpha_1|1\rangle_c + \alpha_2|-1\rangle_c + \alpha_3|0\rangle_c\right) + \\ &|\psi_{11}\rangle_{ab}(\alpha_1\omega^2|1\rangle_c + \alpha_2|-1\rangle_c + \alpha_3\omega|0\rangle_c) + \\ &|\psi_{12}\rangle_{ab}(\alpha_1\omega|1\rangle_c + \alpha_2|-1\rangle_c + \alpha_3\omega^2|0\rangle_c) + \\ &|\psi_{20}\rangle_{ab}(\alpha_1|0\rangle_c + \alpha_2|1\rangle_c + \alpha_3|-1\rangle_c) + \\ &|\psi_{21}\rangle_{ab}(\alpha_1\omega^2|0\rangle_c + \alpha_2|1\rangle_c + \alpha_3\omega|-1\rangle_c) + \\ &|\psi_{22}\rangle_{ab}(\alpha_1\omega|1\rangle_c + \alpha_2|0\rangle_c + \alpha_3\omega^2|-1\rangle_c)\Big\},\end{aligned} \tag{2}$$

where $|\psi_{mn}\rangle = \frac{1}{\sqrt{3}}\sum_{\ell=-1}^{1}\exp(i2n\ell\pi/3)|\ell\rangle|m\ominus\ell\rangle$ forms the complete Bell basis in 3-dimensional OAM subspace, with $m,n \in \{0,1,2\}$ and $\omega = \exp(i2\pi/3)$. Alice routinely needs to distinguish these 9 Bell states by performing HDBSM on photons $a$ and $b$, and then communicates her measurement results to Bob via a classical channel. Accordingly, Bob conducts the desired unitary operations on photon $c$ to recover the original state, and thus accomplishing the teleportation task reliably.

As was well known, HDBSM remains technically challenging in linear optical schemes. Instead, we adopt the nonlinear optical process of SFG to measure the complete Bell basis. By a careful analysis on the OAM conversion behaviors in SFG process, we find that the desired 9 Bell states can all be identified equivalently. First,

Alice can mix photons *a* and *b* directly in the nonlinear crystal. Owing to OAM conservation in the process of SFG[38], we can map the following 7 two-photon OAM Bell states onto 7 single-photon OAM superposition states, respectively, that is,

$$\begin{aligned}
|\psi_{00}\rangle_{ab} &\to |0\rangle, \\
|\psi_{10}\rangle_{ab} &\to |-2\rangle + 2|1\rangle, \\
|\psi_{11}\rangle_{ab} &\to \omega^2|-2\rangle + |1\rangle + \omega|1\rangle, \\
|\psi_{12}\rangle_{ab} &\to \omega|-2\rangle + |1\rangle + \omega^2|1\rangle, \\
|\psi_{20}\rangle_{ab} &\to |2\rangle + 2|-1\rangle, \\
|\psi_{21}\rangle_{ab} &\to \omega^2|-1\rangle + |-1\rangle + \omega|2\rangle, \\
|\psi_{22}\rangle_{ab} &\to \omega|-1\rangle + |-1\rangle + \omega^2|2\rangle.
\end{aligned} \qquad (3)$$

Of particular importance is the well-established mutual orthogonality for these 7 single-photon OAM states at the right-handed sides of Eq. (3). While for $|\psi_{01}\rangle_{ab}$ and $|\psi_{02}\rangle_{ab}$, the SFG for photons *a* and *b* is forbidden, as a result of the destructive interference in SFG light fields. Still, we can identify $|\psi_{01}\rangle_{ab}$ and $|\psi_{02}\rangle_{ab}$ by performing a prior OAM-dependent phase shifts on photon *a*. This can be done simply by inserting into photon *a*'s path one Dove prism (DP)[39], whose orientation is adjusted at an angle of $-\pi/3$ and $\pi/3$, respectively, to convert $|\psi_{01}\rangle_{ab}$ and $|\psi_{02}\rangle_{ab}$ to $|\psi_{00}\rangle_{ab}$. Note that, in addition to preparing the OAM maximal entanglement, the use of perfect vortices is also essential for the faithful SFG-based HDBSM, as it guarantees the identical conversion efficiency for input different OAM modes. In other words, the deterministic high-dimensional teleportation could be realized with our scheme of perfect-vortex-encoded and SFG-based faithful HDBSM in the high-dimensional OAM subspace.

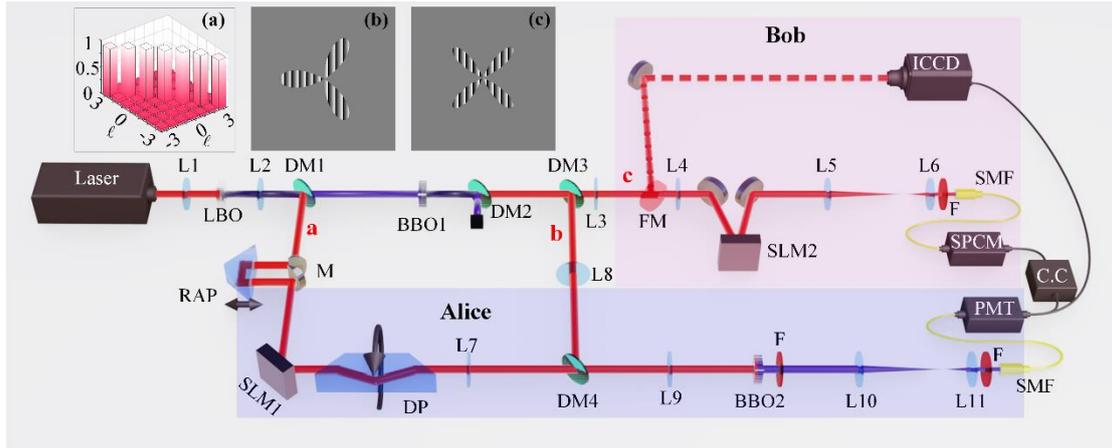

**Fig. 1 | Experimental setup for quantum teleportation of high-dimensional OAM states and optical images.** Inset (a) shows the measured two-photon 7-dimensional OAM spectrum, (b) and (c) illustrate the holographic gratings used for encoding the three-leaf Clover and four-leaf Clover images to be teleported. By switching the flipper mirror (FM), full tomography for the teleported OAM superposition states and the capture of the teleported images can be conducted by using the coincidence circuit (C.C) and the ICCD camera, respectively. See the **Methods** for details.

Figure 1 illustrate our experimental setup for high-dimensional OAM teleportation and for quantum image teleportation. The light beam derived from a mode-locked Ti-sapphire laser has a central wavelength of 710 nm and a duration of 140 fs. We first direct the laser beam to the Lithium triborate crystal (LBO), via second-harmonic generation (SHG), to generate the ultraviolet pulses centered at 355 nm. The residual 710nm light is guided onto a spatial light modulator (SLM1) that displays suitable holograms for preparing photon *a* in the desired high-dimensional OAM superpositions to be teleported. While the generated 355 nm ultraviolet pulses are directed to pump the β-barium borate crystal (BBO1) to create the non-degenerate OAM-entangled photon pairs via spontaneous parametric down-conversion (SPDC), i.e., photon *b* and *c*, centered at 780 nm and 650 nm, respectively. Alice possesses photon *a* and *b* while Bob holds photon *c*. In order to teleport the high-dimensional OAM states encoded by photon *a* onto photon *c*, Alice performs the HDBSM on her photon *a* and *b*, by sending them together to another BBO2 to do SFG. The 372nm SFG photon is then coupled to a single-mode fiber (SMF) and detected by a Photo Multiplier Tube (PMT). As mentioned above, Alice is able to identify all 9 Bell states by prior suitable OAM manipulations and state filtering. Thus the single-photon event from PMT at Alice's side indicates the faithful HDBSM and heralds a teleported photon *c* at Bob's side.

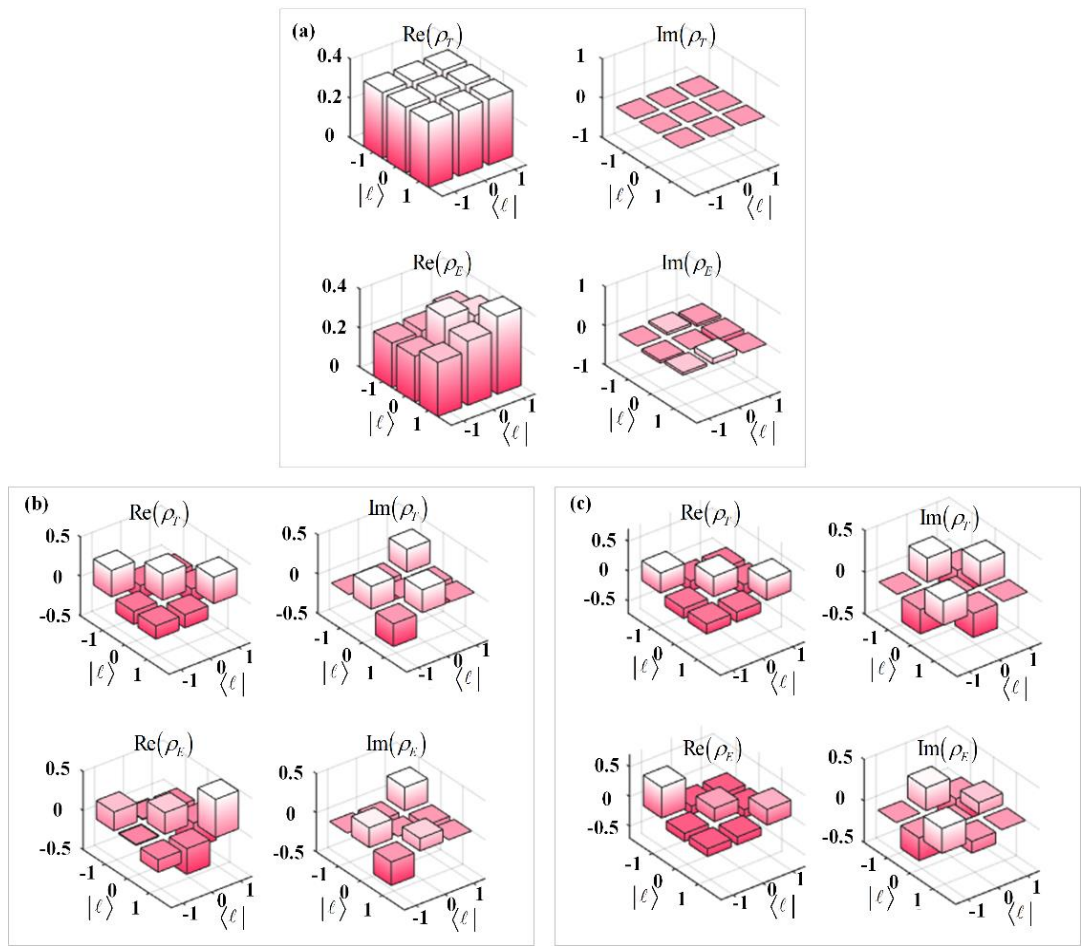

**Fig. 2 | Density matrices for 3-dimensional OAM teleportation,** $|\varphi_4\rangle_a = (|-1\rangle_a + |0\rangle_a + |1\rangle_a)/\sqrt{3}$, **conditional on three specific HDBSMs:** (a) $|\psi_{00}\rangle_{ab}$, (b) $|\psi_{01}\rangle_{ab}$ and (c) $|\psi_{02}\rangle_{ab}$. In each case, the upper rows are the original real and imaginary parts while the bottom are the experimental results.

In our first set of experiment, we prepare 12 OAM states to be teleported at Alice's side, $|\varphi_1\rangle_a = |-1\rangle_a, |\varphi_2\rangle_a = |0\rangle_a, |\varphi_3\rangle_a = |1\rangle_a$, $|\varphi_4\rangle_a = (|-1\rangle_a + |0\rangle_a + |1\rangle_a)/\sqrt{3}$, $|\varphi_5\rangle_a = (|-1\rangle_a + \omega|0\rangle_a + \omega^2|1\rangle_a)/\sqrt{3}$, $|\varphi_6\rangle_a = (|-1\rangle_a + \omega^2|0\rangle_a + \omega|1\rangle_a)/\sqrt{3}$, $|\varphi_7\rangle_a = (\omega|-1\rangle_a + |0\rangle_a + |1\rangle_a)/\sqrt{3}$, $|\varphi_8\rangle_a = (|-1\rangle_a + \omega|0\rangle_a + |1\rangle_a)/\sqrt{3}$, $|\varphi_9\rangle_a = (|-1\rangle_a + |0\rangle_a + \omega|1\rangle_a)/\sqrt{3}$, $|\varphi_{10}\rangle_a = (\omega^2|-1\rangle_a + |0\rangle_a + |1\rangle_a)/\sqrt{3}$, $|\varphi_{11}\rangle_a = (|-1\rangle_a + \omega^2|0\rangle_a + |1\rangle_a)/\sqrt{3}$, and $|\varphi_{12}\rangle_a = (|-1\rangle_a + |0\rangle_a + \omega^2|1\rangle_a)/\sqrt{3}$. They constitute the full four sets of 3-dimensional mutually unbiased bases (MUB), which are commonly used to testify the universality of teleportation for all possible superpositions states[40]. The verification of teleportation results is based on the two-fold coincidence detections between the SFG photon (detected by PMT) and photon $c$ (detected by SPCM). After teleportation, we employ the generalized Gell-Mann matrix basis[41] to reconstruct the density matrices of photon $c$ at Bob's side. Then the effectiveness of our high-dimensional teleportation can be confirmed quantitatively by calculating the teleportation fidelity, $F = \text{Tr}(\rho_E \rho_T) + \sqrt{1 - \text{Tr}(\rho_E^2)}\sqrt{1 - \text{Tr}(\rho_T^2)}$, where $\rho_E$ and $\rho_T$ represent the density matrices of the experimentally measured states and the original ones, respectively. We present in Fig. 2 the teleportation results for a qutrit OAM state, $|\varphi_4\rangle_a = (|-1\rangle_a + |0\rangle_a + |1\rangle_a)/\sqrt{3}$, without loss of generality, conditional on three different specific HDBSM. Specifically, Figure 2(a), 2(b), and 2(c) show the resultant density matrices conditional on HDBSM results of $|\psi_{00}\rangle_{ab}$, $|\psi_{01}\rangle_{ab}$, and $|\psi_{02}\rangle_{ab}$, respectively, in each of which the upper rows are the original real and imaginary parts while the bottom are the experimentally reconstructed ones. The fairly good agreement between them can be seen clearly. For quantitative analysis,

we also show in Fig. 3 the teleportation fidelities for all 12 MUB states conditional on these three HDBSM, all of which are significantly higher than the nonclassical teleportation bound[42] of 1/2. The average teleportation fidelity is $0.8788\pm0.048$, with 7.89 standard deviations above the classical limit 0.5. Note that all these experimental results are obtained without any background subtraction. Besides, the imperfect spatio-temporal overlap of photons $a$ and $b$ at BBO2 is the main limited factor for a better teleportation performance.

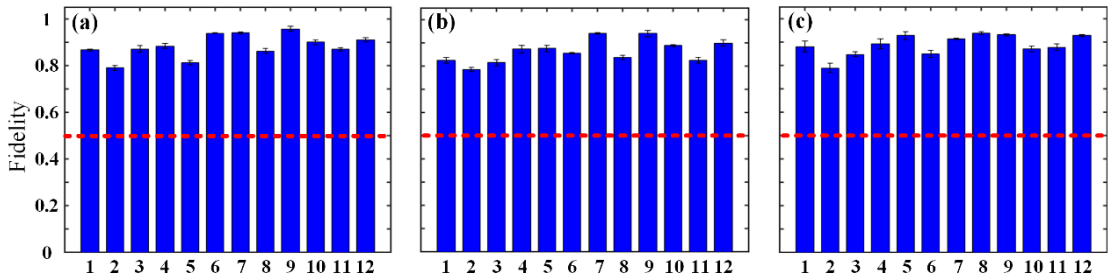

**Fig. 3 | Experimental teleportation fidelities for all 12 OAM MUB states condition on three specific HDBSM results:** (a) $|\psi_{00}\rangle_{ab}$, (b) $|\psi_{01}\rangle_{ab}$, and (c) $|\psi_{02}\rangle_{ab}$. The labels of 1-12 in the horizontal axis represent the OAM qutrit states, $|\varphi_1\rangle_a - |\varphi_{12}\rangle_a$. The red dashed lines indicate the nonclassical quantum teleportation bound 1/2.

Our quantum teleportation scheme for the above OAM qutrit states can be straightforwardly extended to embrace any higher-dimensional states. As is well known, any 2D optical image can be equivalently represented by a high-dimensional state vector encoded in the photonic spatial degrees of freedom, e.g., according to the digital spiral imaging concept[43,44]. In this regard, quantum image teleportation can be reckoned as the ultimate goal of high-dimensional teleportation. Besides, the vision or image is by far the most universal carrier for transferring information in the real world. Therefore, quantum image teleportation is also highly desirable for future high-capacity complex quantum network. For such a quantum image teleporter, Alice and Bob share photon pairs that are entangled in the full transverse spatial modes, which can be written in terms of LG modes as[37], $|\Psi\rangle_{bc} = \sum_{\ell,p} C_{p,p}^{\ell,-\ell} |\ell,p\rangle_b |-\ell,p\rangle_c$, where $C_{p,p}^{\ell,-\ell}$ denotes the probability amplitude of finding photon $b$ in the mode of $|\ell,p\rangle$ and the photon $c$ in the mode of $|-\ell,p\rangle$. The full transverse spatial mode entanglement has been explored for full-field quantum imaging and quantum pattern recognition[45]. However, it has not yet been explored for quantum image teleportation. For this, at Alice's side, the real-amplitude images, e.g., three-leaf and four-leaf clover, are taken as examples for such a purpose. They are prepared and encoded by photon $a$ with the holograms displayed by SLM1, see the insets of Fig. 1(b) and 1(c). Similarly, Alice performs the HDBSM by mixing photons $a$ (bearing the image information) and photon $b$ in the BBO2 crystal to conduct SFG, and then filtering out the fundamental mode of SFG photons using a SMF that is connected to a PMT. The single-photon events from PMT indicates the HDBSM result of $|\psi_{00}\rangle_{ab}$ at Alice's side, which serves the trigger signal for the ICCD camera at Bob's side to capture the image that is teleported onto photon $c$. We present our experimental observations of the teleported three-leaf and four-leaf Clover images in Fig. 4(a) and 4(b), respectively. As shown by the dotted lines, we define the regions of interest (ROI) for both three-leaf and four-leaf Clover images, which occupy about 432 and 441 pixels, respectively. We can define the image fidelity as $F = \left|\sum_{k,j} I_E(k,j) I_O(k,j)\right|^2$, where $I_E(k,j)$ and $I_O(k,j)$ is the normalized amplitude at the pixel position $(k,j)$ for the experimentally teleported image and the

original one, respectively. Accordingly, we have $F = 0.5536$ and 0.4861 for three-leaf and four-leaf Clover images, respectively. We can attribute the low fidelity of quantum image teleportation mainly to both the non-maximal spatial entanglement for photon pairs generated by SPDC and the nonuniform SFG efficiency for each pixel in the HDBSM stage.

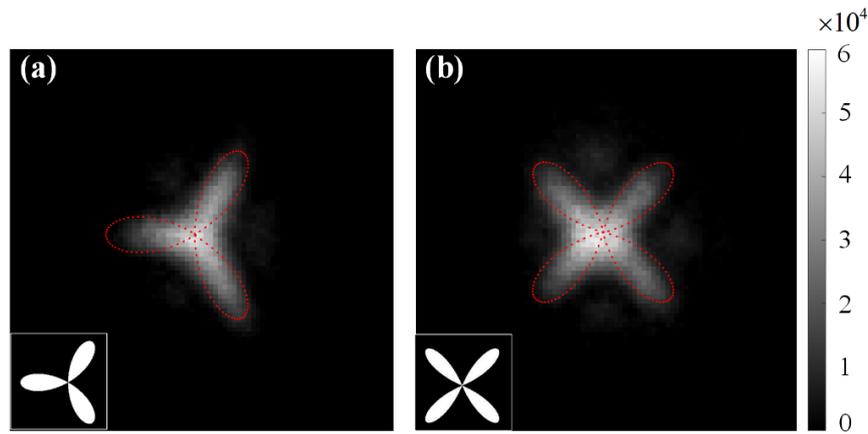

**Fig. 4 | Experimental results for quantum image teleportation:** (a) Three-leaf Clover and (b) four-leaf Clover. Insets show the original images. The gray-scale bar is in units of photons/pixel.

In our scheme, the teleported images are recorded and visualized directly with a spatially resolved ICCD camera, i.e., pixelated. Each pixel can be considered as a single spatial mode individually. Therefore it is illuminating for us to characterize the teleportation channel capacity from a multi-pixel field of view. In theory, the amplitude or wave function of the teleported image can be equivalently expressed by, $|\varphi\rangle_{image} = \sum_{k,j} f(k,j) |k,j\rangle$, where the computational basis $|k,j\rangle$ encodes the pixel position $(k,j)$, and $f(k,j)$ represents the normalized pixel grayscale value, with $\sum_{k,j} |f(k,j)|^2 = 1$. A successful teleportation of an image means that individual pixel information should be teleported successfully. For two-photon maximal entanglement and a perfect HDBSM, the successful teleportation can be manifested by the teleported image that is identical to the original one. In other words, the pixelated images can offer an intuitive way to characterize the effective dimensionality of the nonlocal channels that sustain the quantum image teleportation, in which the Schmidt number[46] can be written as, $\kappa = 1/\sum_{k,j}|f(k,j)|^4$. Accordingly, from the experimentally recorded images by the ICCD camera, we can directly calculate the Schmidt numbers as $\kappa = 134.68$ and 106.33, thus showcasing the high-dimensional feature of our quantum image teleporter systems.

In summary, we have demonstrated quantum teleportation of high-dimensional spatial modes and even a quantum image teleporter. The key to our scheme is the exploitation of perfect vortices both for making two-photon high-dimensional yet maximal OAM entanglement and for performing a high-effective yet faithful HDBSM via SFG. We have experimentally achieved the average teleportation fidelity of $0.8788 \pm 0.048$ for a complete set of 3-dimensional MUB, significantly beyond the classical limit. More importantly, we succeed in realizing the first quantum image teleportation by exploring the full transverse spatial entanglement. We use the ICCD camera to capture the teleported images, and from the multi-pixel field of view, we estimate the effective channel capacity $\kappa > 100$, thus showcasing the inherent high-dimensional feature of our quantum image teleportation system. Our scheme is fundamentally different previous quantum imaging techniques[47], such as ghost imaging[48] or quantum imaging with undetected photons[49]. Our method holds promise for future high-volume and secure quantum image transmission.

**References**

1.  Bennett, C. H., Brassard, G., Créepeau, C., Jozsa, R., Peres, A. & Wootters, W. K. Teleporting an unknown quantum state via dual classical and Einstein-Podolsky-Rosen Channels. *Phys. Rev. lett.* **70**, 1895-1899 (1993).

2.  Bouwmeester, D. et al. Experimental quantum teleportation. *Nature* **390**, 575–579 (1997).

3.  Marcikic, I., de Riedmatten, H., Tittel, W., Zbinden, H. & Gisin, N. Long-distance teleportation of qubits at telecommunication wavelengths. *Nature* **421**, 509–513 (2003).

4.  Takeda, S., Mizuta, T., Fuwa, M., Loock, P. & Furusawa A. Deterministic quantum teleportation of photonic quantum bits by a hybrid technique. *Nature* **500**, 315-318 (2013).

5.  Boschi, D., Branca, S., De Martini, F., Hardy, L. & Popescu, S. Experimental realization of teleporting an unknown pure quantum state via dual classical and Einstein–Podolsky–Rosen channels. *Phys. Rev. Lett.* **80**, 1121–1125 (1998).

6.  Furusawa, A. et al. Unconditional quantum teleportation. *Science* **282**, 706–709 (1998).

7.  Nielsen, M. A., Knill, E. & Laflamme, R. Complete quantum teleportation using nuclear magnetic resonance. *Nature* **396**, 52–55 (1998).

8.  Kim, Y., Kulik, S. & Shih, Y. Quantum Teleportation of a Polarization State with a Complete Bell State Measurement. *Phys. Rev. Lett.* **86**, 1370-1373 (2001).

9.  Ursin, R. et al. Communications: quantum teleportation across the Danube. *Nature* **430**, 849–849 (2004).

10. Barrett, M. et al. Deterministic quantum teleportation of atomic qubits. *Nature* **429**, 737–739 (2004).

11. Riebe, M. et al. Deterministic quantum teleportation with atoms. *Nature* **429**, 734–737 (2004).

12. Sherson, J. F. et al. Quantum teleportation between light and matter. *Nature* **443**, 557–560 (2006).

13. Yonezawa, H., Aoki, T. & Furusawa A. Demonstration of a quantum teleportation network for continuous variables. *Nature* **431**, 430-433 (2004).

14. de Riedmatten, H. et al. Long distance quantum teleportation in a quantum relay configuration. *Phys. Rev. Lett.* **92**, 047904 (2004).

15. Olmschenk, S. et al. Quantum Teleportation Between Distant Matter Qubits. *Science* **323**, 486-

489 (2009).

16. Yin, J. et al. Quantum teleportation and entanglement distribution over 100-kilometre free-space channels. *Nature* **488**, 185–188 (2012).

17. Ma, X.-S. et al. Quantum teleportation over 143 kilometres using active feed-forward. *Nature* **489**, 269–273 (2012).

18. Krauter, H. et al. Deterministic quantum teleportation between distant atomic objects. *Nat. phys.* **9**, 400-404 (2013).

19. Pfaff, W. et al. Unconditional quantum teleportation between distant solid-state quantum bits. *Science* **345**, 532–535 (2014).

20. Ren, J. et al. Ground-to-satellite quantum teleportation. *Nature* **549**, 70-73 (2017).

21. Wang, X.-L. et al. Quantum teleportation of multiple degrees of freedom of a single photon. *Nature* **518**, 516–519 (2015).

22. Luo, Y. et al. Quantum Teleportation in High Dimensions. *Phys. Rev. Lett.* **123**, 070505 (2019).

23. Hu, X.-M. et al. Experimental High-Dimensional Quantum Teleportation. *Phys. Rev. Lett.* **125**, 230501 (2020).

24. Chen. L.-X. Quantum discord of thermal two-photon orbital angular momentum state: mimicking teleportation to transmit an image. *Light Sci. Appl.* **10**, 148 (2021)

25. Erhard, M., Krenn, M. & Zeilinger, A. Advances in high-dimensional quantum entanglement. *Nat. Rev. Phys.* **2**, 365–381 (2020).

26. Ostrovsky, A. S., Rickenstorff-Parrao, C. & Arrizón, V. Generation of the "perfect" optical vortex using a liquid-crystal spatial light modulator. *Opt. Lett.* **38**, 534–536 (2013)..

27. Briegel, H. J., Dur, W., Cirac, J. I. & Zoller, P. Quantum repeaters: the role of imperfect local operations in quantum communication. *Phys. Rev. Lett.* **81**, 5932–5935 (1998).

28. Kimble, H. J. The quantum internet. *Nature* **453**, 1023–1030 (2008).

29. Gottesman, D. & Chuang, I. Demonstrating the viability of universal quantum computation using teleportation and single-qubit operations. *Nature* **402**, 390–393 (1999).

30. Knill, E., Laflamme, R. & Milburn, G. J. A scheme for efficient quantum computation with linear optics. *Nature* **409**, 46–52 (2001).

31. Bechmann-Pasquinucci, H., & Tittel, W. Quantum cryptography using larger alphabets. *Phys. Rev. A* **61**, 062308 (2000).

32. Cerf, N. J., Bourennane, M., Karlsson, A. & Gisin, N. Security of Quantum Key Distribution Using d-Level Systems. *Phys. Rev. Lett.* **88**, 127902 (2002).

33. Lanyon, B. P. et al. Simplifying quantum logic using higher-dimensional Hilbert spaces. *Nat. Phys.* **5**, 134-140 (2009).

34. Son, W., Lee, J., Kim, M. & Park, Y.-J. Conclusive teleportation of a d-dimensional unknown state. *Phys. Rev. A* **64**, 064304 (2001).

35. Padgett, M. J. Orbital angular momentum 25 years on [Invited]. *Opt. Express* **25**, 11265-11274 (2017)

36. Allen, L., Beijersbergen, M. W., Spreeuw, R. J. C. & Woerdman, J. P. Orbital angular momentum of light and the transformation of Laguerre-Gaussian laser modes. *Phys. Rev. A* **45**, 8185 (1992).

37. Torres, J. P., Alexandrescu, A. & Torner, L. Quantum spiral bandwidth of entangled two-photon states. *Phys. Rev. A* **68**, 050301(R) (2003).

38. Qiu, X.-D. et al. Optical vortex copier and regenerator in the Fourier domain. *Photonics Research* **6**, 641-646 (2018).

39. Wang. F. Generation of the complete four-dimensional Bell basis. *Optica* **4**, 1462-1467 (2017).

40. Ivanovic, I. D. Geometrical description of quantal state determination. *J. Phys. A*: *Math. Gen.* **14**, 3241-3245 (1981).

41. Bertlmann, R. A. & Krammer, P. Bloch vectors for qudits. *J. Phys. A: Math. Theor.* **41**, 235303 (2008).

42. Hayashi, A., Hashimoto, T. & Horibe, M. Reexamination of optimal quantum state estimation of pure states. *Phys. Rev. A* **72**, 032325 (2005).

43. Torner, L. Torres, J. P. & Carrasco, S. Digital spiral imaging. *Opt. Exp.* **13**, 873-881 (2005).

44. Chen. L., Lei, J. & Romero, J. Quantum digital spiral imaging. *Light Sci. Appl.* **3**, e153 (2018).

45. Qiu, X., Zhang, D., Zhang, W. & Chen, L. Structured-Pump-Enabled Quantum Pattern Recognition. *Phys. Rev. Lett.* **122**, 123901 (2019).

46. Law, C. K. & Eberly, J. H. Analysis and interpretation of high transverse entanglement in optical parametric down conversion. *Phys. Rev. Lett.* **92**, 127903 (2004).

47. Moreau, P. A., Toninelli, E., Gregory, T. & Padgett, M. J. Imaging with quantum states of light. *Nat. Rev. Phys.* **1**, 367-380 (2019).

48. Pittman, T. B., Shih, Y. H., Strekalov, D. V. & Sergienko, A. V. Optical imaging by means of two-photon quantum entanglement. *Phys. Rev. A* **52**, R3429-R3432 (1995).

49. Lemos, G. B. et al. Quantum imaging with undetected photons. *Nature* **512**, 409-412 (2014).

## Methods

**Perfect vortices.** Photon pairs generated by SPDC have proven to be a reliable entanglement source. However, under the thin-crystal approximation and phase-matching condition with a Gaussian pump beam, the down-converted two-photon OAM entanglement inevitably suffers from the limited spiral bandwidth[37], i.e., lower-order LG modes appears more frequently than higher-order ones. If the standard LG modes are used to represent the OAM eigenstates $|\ell\rangle$, namely, $\langle r,\phi|\ell\rangle = \mathrm{LG}_{p=0}^{\ell}(r,\phi)$, the two-photon OAM entangled state can be written as, $|\Psi\rangle_{bc} = \sum_{\ell} C_{\ell} |\ell\rangle_b |-\ell\rangle_c$, where $C_{\ell} = \int \mathrm{LG}_0^0(r,\phi)\left[\mathrm{LG}_0^{\ell}(r,\phi)\right]^* \left[\mathrm{LG}_0^{-\ell}(r,\phi)\right]^*$ represents the probability amplitude of finding photon $b$ in the mode of $|\ell\rangle$ and the photon $c$ in the mode of $|-\ell\rangle$. As mentioned above, $|\Psi\rangle_{bc}$ is merely a non-maximally entangled OAM state. However, the maximal entanglement is prerequisite for a standard quantum teleportation. To overcome this obstacle, we adopt the so-called "perfect vortices[26]" to represent OAM eigenstate, instead of the standard LG modes. In our scheme, we prepare the perfect vortex states by modifying the LG modes as[50],

$$\mathrm{MLG}_0^{\ell}(r,\phi) = \left|\mathrm{LG}_0^l(r,\phi)\right| \exp(i\ell\phi), \qquad (4)$$

where $l$ is a constant, e.g., $l=5$. In such a basis of perfect vortices, we know that $C_{\ell} = \int \mathrm{LG}_0^0(r,\phi)\left[\mathrm{MLG}_0^{\ell}(r,\phi)\right]^* \left[\mathrm{MLG}_0^{-\ell}(r,\phi)\right]^*$ will become a constant, as they share the same intensity distribution and have the identical overlap probability. Thus we can obtain the desired maximally entangled OAM state as, $|\Psi\rangle_{bc} = \frac{1}{\sqrt{d}}\sum_{\ell}|\ell\rangle_b|-\ell\rangle_c$, with $d$ being the dimensionality of the OAM subspace.

Besides, the perfect vortex is also crucial for performing the faithful HDBSM via SFG. Based on the couple-wave equations describing the SFG, we can estimate the frequency conversion efficiency for our perfect vortices of modified LG mode as[51],

$$\eta_{\ell_a,\ell_b} \propto \int \text{MLG}_0^{\ell_a}(r,\varphi)\text{MLG}_0^{\ell_b}(r,\varphi)[\text{MLG}_0^{\ell_a+\ell_b}(r,\varphi)]^* r dr d\varphi. \tag{5}$$

Similarly, we can expect that $\eta_{\ell_a,\ell_b}$ will become a constant even for different OAM modes, and thus realizing a faithful HDBSM.

**Experimental Setup.** Our experimental setup for high-dimensional teleportation is sketched in Fig. 1. The 140 fs pulsed laser beam with an average power of 1950 mW, a central wavelength of 710 nm and a repetition rate of 80 MHz is focused by lens L1 ($f_1 = 150$mm) and guided to pump a 1.5-mm-long LBO crystal to generate the ultraviolet pulse of a central wavelength of 355 nm and an average power of 200 mW via second-harmonic generation (SHG). The residual 710nm light beam is collimated by L2 ($f_2 = 100$mm) and guided onto a spatial light modulator (SLM1), which is used to display suitable holographic gratings for preparing photon $a$ in an arbitrary high-dimensional OAM superposition state to be teleported. After reflection from SLM1, the light beam has a power of roughly 140 mW. While the generated 355 nm SHG ultraviolet pulses are directed to pump a 3-mm-long β-barium borate crystal (BBO1) to create the frequency non-degeneracy OAM entangled photon pairs via SPDC, i.e., photon *b* and *c*, whose wavelengths are centered at 780 nm and 650 nm, respectively. A long-pass dichroic mirror (DM2) is inserted to separate the down-converted photons from the pump. For performing HDBSM, the 780 nm photon *b* reflected from DM3 and photon *a* are combined through DM4, and then sent to 1.5-mm-long BBO2 for implementing SFG. Note that, photon *a* and *b* are both imaged onto the BBO2 via 4*f* imaging systems, L7-L9 and L8-L9 ($f_7 = f_8 = 500\text{mm}$, $f_9 = 100\text{mm}$), to ensure a good spatial overlap of the image planes of BBO2, BBO1 and SLM1. In addition, the temporal overlap between photon *a* and *b* in BBO2 is realized by accurately adjusting the position of right-angle prism (RAP). Subsequently, we use another 4*f* system ($f_{10} = 500\text{mm}$ and $f_{11} = 300$) together with a 4 mm collimated lens to couple the generated 372nm SFG photon into a single-mode fiber (SMF), which is connected to a Photo Multiplier Tube (PMT). The single-photon event from PMT indicates the successful BSM signal and herald a teleported photon *c* at Bob's side. For performing the state tomography, photon *c* is imaged onto SLM2 with a 4*f* imaging system ($f_3 = 200\text{mm}$ and $f_4 = 400\text{mm}$) and then reimaged onto the input facet of SMF with another 4*f* imaging system ($f_5 = 500\text{mm}$ and $f_6 = 10\text{mm}$). By switching the flipper mirror (FM), we can capture the teleported images by using the ICCD camera, which is triggered by the single-photon events from the PMT. Note that an image preserving optical delay of about 25m is requested to compensate the electronic delay associated with the PMT and the trigger mechanism in the ICCD camera[45].

50. Chen, L., Zhang, W., Lu, Q. & Lin, X. Making and identifying optical superpositions of high orbital angular momenta. *Phys. Rev. A* **88**, 053831 (2013).

51. Zhou, Z. Y. et al. Orbital angular momentum photonic quantum interface. *Light: Science & Applications* *5*, e16019-e16019 (2016).